# Benefits and Challenges of Dynamic Modelling of Cascading Failures in Power Systems

Yitian Dai[1], Robin Preece[1], and Mathaios Panteli[2]
[1]Department of Electrical and Electronic Engineering, The University of Manchester, Manchester, UK
[2]Department of Electrical and Computer Engineering, University of Cyprus, Nicosia, Cyprus
{yitian.dai; robin.preece}@manchester.ac.uk
panteli.mathaios@ucy.ac.cy

*Abstract*—Time-based dynamic models of cascading failures have been recognized as one of the most comprehensive methods of representing detailed cascading information and are often used for benchmarking and validation. This paper provides an overview of the progress in the field of dynamic analysis of cascading failures in power systems and outlines the benefits and challenges of dynamic simulations in future grids. The benefits include the ability to capture temporal characteristics of system dynamics and provide timing information to facilitate control actions for blackout mitigation. The greatest barriers to dynamic modelling of cascading failures are the computational burden, and the extensive but often unavailable data requirements for dynamic representation of a power system. These factors are discussed in detail in this paper and the need for in-depth research into dynamic modelling of cascading failures is highlighted. Furthermore, case studies of dynamic cascading simulation of 200-bus and 2000-bus benchmark systems provide initial guidance for the selection of critical parameters to enhance simulation efficiency. Finally, cross-validation and comparison against a quasi-steady state DC power flow model is performed, with various metrics compared.

*Index Terms*--Cascading failures, dynamic power system simulation, frequency stability, power system reliability.

## I. Introduction

Power systems are undergoing a period of rapid evolution towards more intelligent and sustainable smart grids. While improving economic benefits and operational efficiency, emerging technologies have brought new challenges to the continuous and secure operation of power systems. Recent blackouts, e.g., the 2015 blackout in Turkey [1] and the 2018 blackout in Canada [2], have raised concerns about power system reliability and have highlighted the importance of developing methods aimed at evaluating and mitigating the risks associated with blackouts.

Cascading failures in power systems can be defined as "the uncontrolled successive loss of system elements triggered by an incident at any location" [3]. They are usually triggered by one or more unplanned disturbance caused by component aging, bad weather, poor maintenance or maloperation, resulting in thermal overload and/or dynamic instability of other components, thus starting the cascading failures [4]. The propagation of cascading failures is governed by the details of the system operating state, including power flow distribution, frequency excursions, voltage profile, control, and protection actions. The complexity of power system dynamics necessitates extensive simulation work to reproduce system behaviours, establish the causal sequence of cascading events and develop feasible control strategies. Hence, there is a growing need for in-depth understanding of the complex dynamics involved in cascading failure propagation to address this analytical need.

Analysing and resolving cascading failures, one of the main causes of large blackouts, has been considered an important aspect of improving system performance and facilitating decision making [4]–[6]. Doing so requires detailed modelling of system components that can capture system behaviours during cascades and provide accurate and useful information for cascading risk management. Existing cascading failure models, incorporating electrical properties, can be roughly grouped into three categories: 1) quasi-steady state (QSS) models, which simplify continuous dynamics to a series of discrete state transitions [7], 2) dynamic models, which explicitly simulate transient dynamics following each event [8], [9], and 3) quasi-dynamic models, which combine QSS models and dynamic models, and apply them on different timescales as desired [10]. The use of time-based dynamic models has been recognised as the one of the most comprehensive methods to reveal how failures propagate [8]. The dynamic models can be computationally expensive for large-scale statistical studies but comparing a comprehensive model with a simplified model can be useful in understanding the validity of different model simplifications to support benchmarking. Moreover, a dynamic model is indispensable for assessing mitigation techniques, which require detailed system dynamics to thoroughly test their capabilities.

Several issues make the development of *dynamic* cascading failure models increasingly important and challenging with respect to understanding, simulating and mitigating cascading failures in evolving power grids. Particularly, the larger problem scale, the need to simulate more variables describing control and protection actions, the more complicated system fre-quency characteristics due to low inertia issues, and the increased voltage instability issues due to the increased proportion of power electronics-based generation. The potential challenges of large-scale renewable energy integration on system inertia, voltage stability, and more importantly, transient phenomena have attracted the interest of some researchers, such as [11]–[13]. This necessitates a time-based dynamic model. However, the impact of increased renewable penetration on cascading failure propagation has received less attention. Thus motivated, the aims of this paper are: 1) to emphasize the need for intensive research effort in dynamic modelling of cascading failures in evolving power grids, 2) to present an overview of research associated with dynamic modelling of cascading failures, 3) to discuss the challenges arising from dynamic simulation and promising solutions to help overcome these limitations, and 4) to provide initial guidance on the selection of critical parameters to improve dynamic cascading failure simulation efficiency, as well as methods for model validation.

The rest of this paper proceeds as follows: Section II describes the need for dynamic modelling of cascading failures. Section III first introduces the various power system dynamics and their modelling methods. An overview of various metrics and existing dynamic models is provided, followed by a discussion the challenges of dynamic modelling and possible solutions. Case studies performed using a 200-bus and a 2000-bus system are carried out to illustrate the selection of parameters in Section IV. Following this, a comparative study between a dynamic model and a QSS DC power flow model is presented. Section V concludes the paper.

II. NEED FOR DYNAMIC MODELLING OF CASCADING FAILURES

Cascading failure modelling in power systems generally considers a set of cascading mechanisms and dynamic instability phenomena, including thermal overload of transmission lines, protection relay operation, frequency control, voltage collapse, inter-area oscillations and other mechanisms identified by IEEE and CIGRE Working Groups on Cascading Failure [6], [14]. Furthermore, historical blackouts have also shown the important roles of communication systems and human activities, such as during the 2015 Ukraine blackout [15]. Various dynamics exhibit distinct timescales of operation, so only with an appropriate representation of time, can the model reflect interactions between relevant mechanisms and provide reliable conclusions. Moreover, application of time-based models can potentially lead to improvement in system performance with respect to cascading events, such as evaluating and optimizing the time settings of control and protection actions. Therefore, it is important to consider the temporal characteristics of system components when performing cascading failure analysis.

In existing literature, cascading failures are mainly analysed using QSS models based on AC/DC power flow. QSS models have the advantages of being computationally efficient and able to obtain the representative properties of cascading failures, e.g., self-organized criticality [16]. Based on the linear power flow assumption, the QSS DC power flow model is widely used, however it fails to capture important aspects in evolving power systems, such as voltage behaviour and reactive power flows, which have been shown to have significant effects in past power outages [17]. The QSS AC power flow considers non-linear phenomena, but requires assumptions to specify generator re-dispatch capabilities, load shedding capabilities and to handle non-convergent conditions [18]. At present, there is no clear consensus on the details of modelling assumptions for cascading failure simulation and analysis, as they vary in different systems and under different operating conditions [19]. This poses challenges for benchmarking and validation of different cascading failure models. Therefore, there is a need to develop a generic dynamic simulation platform to address the requirements of cascading failure analysis in evolving power systems. On this basis, tightly temporally coupled mechanisms can be rigorously investigated, and advances can be made in understanding their relative importance and determining the level of modelling detail required to address specific research questions.

In response to the raising awareness of the energy crisis and climate change, power systems are undergoing several changes that may increase the risks associated with cascading blackout:

1) *Increased penetration of renewable energy sources (RES).* As stochastic RES (e.g., wind and solar radiation) continue to replace conventional thermal plants, the growing uncertainty in system operation can increase the risk of cascading blackout concerns.

2) *Growing electrification of modern societies.* The growing electrification of heat and transportation will lead to increased use and reliance on electric power systems and the consequences of electrical system failure will be far larger. Additionally, power systems are becoming increasingly complex and interconnected with greater chance of failure propagations, which could result in more significant detrimental effects of cascading failure.

3) *Widespread deployment of information and communication technologies (ICT).* The extensive deployment of ICTs, as targets of cyber-physical threats, brings new vulnerability and security concerns, e.g., the Ukrainian cyber-attack in 2015 [15].

These emerging changes pose new operating challenges to future renewable-rich, interconnected power grids, and result in complicated system dynamics of which there is currently little understanding. The conventional QSS model assumes that no violations of branch flow or stability limits occur during generator re-dispatch and, therefore, that the power system remains stable before the next event in the cascade occurs. However, with the increasing integration of RES, system inertia is decreasing, and frequency deviations are becoming more severe. This makes the steady-state assumption in the QSS model no longer valid. In this context, the QSS model is no longer sufficient or reliable in characterizing system dynamics during failure propagation. There is a clear necessity to develop time-based dynamic models to accurately capture the transient behaviours in evolving power systems and reflect the nature of cascading outages in a more realistic manner.

III. OVERVIEW OF DYNAMIC ANALYSIS OF CASCADING FAILURE SIMULATORS

Previous research papers that provide overviews of the approaches for cascading failure modelling include [7], [20], [21]. Existing cascading failure models typically simulate a subset of system dynamics according to different research objectives, showing advantages and efficiencies in specific applications, while simplifying in other areas with less impact. The modelling of system dynamics will be discussed in more detail in this section, with insights gained from experiences with past blackouts. Following this, the metrics to quantify cascading effects and existing dynamic cascading simulators are summarized, and the challenges and promising solutions of dynamic cascading simulation are discussed.

### A. Modelling of System Dynamics

The dynamic simulation of evolving power systems, its development and applications have been investigated by several researchers, such as [22]–[24]. A discussion of the infrastructure transitions being implemented in power systems, and the dynamic modelling requirements for conventional and future power system component can be found in [25]. Here, the fundamental elements required to complete studies on cascading failure analysis are shown in Table I, including emerging requirements driven by system evolution. Models of all major components required for dynamic studies in traditional and future power systems are presented and categorized into generation, transmission and distribution networks, consumers, control, and protection devices. Significant infrastructure transitions are being implemented in power systems by the integration of renewable energy sources, the increasing reliance on the electrical energy sector, and the installation of intelligent control techniques such as intentional islanding and coordinated control. These transitions have altered the methods of power system operation and control (e.g., employing bidirectional power transfer and wide-area coordinated control), and pushed the dynamic modelling of cascading failure to new areas. This subsection starts with a brief description of the various controllers and protection devices participated in cascading failure propagation (as summarized in Table I). Then, simulation of future grid technologies such as renewable generators, system-wide control, and demand management is discussed, with an emphasis on emerging requirements for cascading failure analysis. Finally, a generic framework for dynamic cascading failure modelling is provided.

TABLE I
BASIC ELEMENTS REQUIRED FOR CASCADING FAILURE MODELLING IN CONVENTIONAL AND FUTURE POWER GRIDS

|  | Today | Future |
|---|---|---|
| Generation | Synchronous generators | Renewable generation<br>Storage systems |
| Transmission & Distribution | One-way power transfer | Two-way power transfer |
| Consumer | Traditional consumers | Producers and consumers<br>Electric vehicles<br>Smart building |
| Control | Local control: GOV, AGC, AVR, PSS | System-wide control based on wide-area measurements |
| Protection | Overload protection<br>UFLS, OFGT<br>UVLS, OXL, UXL<br>Out-of-step protection | Wide-area coordinated Protection<br>Intentional Islanding |

A synchronous generating unit is generally composed of a prime mover, a synchronous machine, and an excitation system. According to the IEEE guide [26], synchronous generators can be modelled with different levels of complexity, e.g., the standard $6^{th}$ order model and the classical $2^{nd}$ order model. In response to varying reactive power demands, the excitation system keeps the terminal voltage stable by manipulating the field voltage, which is achieved by the automatic voltage regulator (AVR). A power system stabiliser (PSS) can also be fitted to reduce rotor speed oscillations, and the over-/under-excitation limiters (OXL/UXL) are required to protect the machine. If the terminal voltage drops below a critical limit, system-level protection relays such as under-voltage load shedding (UVLS) will act to shed load until an acceptable bus voltage is restored. If one generator loses synchronism with the power system, it will be disconnected by an out-of-step blocking relay. In addition, considering sudden active power mismatch resulting in frequency deviations, the speed governor (GOV) and automatic generation control (AGC) will adjust the generator outputs to re-establish the power balance. For emergency cases where the frequency goes outside the acceptable operating range before reaching a new equilibrium, protection schemes such as under-frequency load shedding (UFLS) and over-frequency generator tripping (OFGT) are used to handle the mismatch. Further, each transmission line is protected by a thermal relay to prevent overloading. An inverse-time delay is usually applied to determine the triggering threshold for the thermal relay.

As summarized in [19], dynamic data of generators and their controllers are available in some power system test cases, e.g., IEEE 39-bus and 68-bus test systems [27], [28], the Illinois 200-bus synthetic system and the Texas 2000-bus synthetic system [29]. Various relay settings can be found in National Grid Codes [30], ENTSO-E [31] and NERC standards [32]. In terms of developing a realistic system model, e.g., for blackout reproduction, measurements of system events are needed for dynamic parameter tuning [33].

Major changes are currently being made to power systems by increasing the penetration of renewable generation, increasing flexible demand, deploying high voltage direct current transmission lines and wide-area controllers, etc. Although system operators require wind generators to be able to ride through certain under-voltage conditions, converter-interfaced RES do not contribute to system inertia unless this behaviour is intentionally designed [4]. Rapid frequency variation can lead to the maloperation of control and protection devices designed for systems with high inertia levels, resulting in a higher risk of cascading failures. Further security concerns arise from voltage stability issues, exacerbated by the replacement of reactive power reserves in thermal power plants with wind generators and static VAR compensators [34]. System-wide coordinated control offers a promising solution to mitigate cascading failures in future grids. Using advanced communication techniques, wide-area monitoring of system global dynamics can potentially prompt the development of coordinated protection and control schemes. Most importantly, these technologies have clear temporal characteristics, and a detailed dynamic model is essential to thoroughly evaluate the extent to which they will alleviate or exacerbate the situation.

Dynamic models compute the effects of initial disturbances by solving differential-algebraic equations (DAE) and monitor the discrete cascading events that may lead to network separation. A generic framework for dynamic cascading failure modelling is specified in Algorithm 1. This algorithm requires, as inputs, network data on topological features, dynamic data of system components and controllers, a power flow solver to compute the initial system state, a simulation step size, a pre-set simulation time, and a set of initial failure scenarios. The selection of these various inputs is discussed further in Section IV-C. Eventually, following the time-stepping simulation including all switching events and dynamic state evolution, a cascading chain of events and total demand loss will be recorded for each failure scenario. Based on this, various cascading metrics can be computed, as discussed in Section III-B.

---

**Algorithm 1:** Dynamic Modelling Algorithm

**Inputs:** Network data, dynamic data of system components and controllers, a power flow solver, a simulation step size, simulation time, initial failure scenario $s \subset S$

**Procedure:**
1: Initialise system state.
2: **for** $s \leftarrow 1, |S|$ **do**
3:     Execute the initial outage(s).
4:     Check for network separation.
       **if** Yes, **then**
           Select a reference machine in each island.
           Update network admittance matrix.
5:     Check for island(s) with no generator.
       **if** Yes, **then**
           Trip the corresponding island(s).
6:     Simulate all islands simultaneously until any of the following occurs:
       **if** Simulation reaches the pre-set time, **then**
           Go to Step 8.
       **if** Any relay threshold is crossed, **then**
           Go to Step 7.
7:     Check for discrete events.
       **if** Yes, **then**
           Execute the event by changing relay output signal to TRUE.
           Go to Step 4.
       **else**
           Go to Step 6.
8: End of simulation.

**Outputs:** A cascading chain of events and total demand loss for each initial failure scenario $s \subset S$

---

### B. Metrics to Quantify Cascading Effects

It is well understood that the resulting impact of cascading failures is difficult to comprehensively assess, as there are long-term potential losses, such as indemnity and maintenance costs caused by power outages. With a focus on power system operation, the impact of a cascading failure is generally quantified by standard metrics, including [8], [19], [35]:

1) *Blackout size*: the amount of unserved demand, in MW or MWh.

2) *Cascade propagation profile*: the total number of line outages, or the number of line outages in each generation/iteration, or the distance between successive disconnected lines.

Many further metrics are defined in the literature to quantify the impact of cascading failures on specific research areas [4], [8], [36], [37]. For example, Expected Demand Not Served (EDNS), Value at Risk (VaR) and Conditional Value at Risk (CVaR) are used to evaluate the blackout size, as defined in (1)-(3). For a given set of failure scenarios ($S$), EDNS provides a mean demand that cannot be met due to cascading failures, where $P(s)$ and $I(s,x)$ are the probability and impact of scenario $s$. The definition of VaR$_\alpha$ given by Choudhry in [38] is "VaR is a measure of market risk. It is the maximum loss which can occur with $\alpha$ % confidence". CVaR$_\alpha$ is closely related to VaR$_\alpha$ and gives the mean loss within $(100 - \alpha)$% confidence. More specifically, VaR$_{95}$ indicates that there is a 95% chance that demand loss will be lower than the specified value, and CVaR$_{95}$ gives the mean unserved demand for the worst 5% cases. Moreover, the empirical probability distribution functions (EPDF) or complementary cumulative distribution functions (CCDF) of demand losses are typically used to verify the simulation results with historical data. The heavy-tailed distribution and power-law fitting are the two key properties of these distributions, which are generally observed in historical data and simulated cascades [39].

$$EDNS(x) = \sum_{\forall s \subset S} P(s)I(s,x) \quad (1)$$

$$VaR_\alpha(x) = min\{x \mid f_x(z) \geq \alpha\} \quad (2)$$

$$CVaR_\alpha(x) = E[x \mid x \geq VaR_\alpha(x)] \quad (3)$$

The cascade propagation profile is usually represented by the probability distribution of the total line outages [8], [19], the number of line outages in each iteration (i.e., quantifying the speed of cascade propagation) [35], [40], [41] and the distance between successive line outages (i.e., quantifying the spatial propagation of cascading outages) [41], [42]. Dobson defined two metrics to quantify the propagation of line outages, i.e., lambda [40] and network distance [41]. The propagation $\lambda_k$ represents the propagation speed of the cascades. As defined in (4), $\lambda_k$ is determined by dividing the mean number of line outages in iteration $k$ by the mean number of line outages in iteration $k-1$, among different cascades. This measure assumes discrete iterations of outages which is not the case when using dynamic time domain simulations, or when analysing historic data. Therefore, successive outages are grouped first into different cascades and then into iterations within each cascade, according to the time interval between them. In [40], outages with a time interval of more than 1 hour belong to different cascades, and outages with a time interval of less than 1 minute are grouped into the same iteration. The network distance quantifies the spatial spreading of the cascading line outages, which can be defined as the minimum number of buses $d^{bus}(L_i, L_j)$ or the minimum length in miles $d^{mile}(L_i, L_j)$ in a network path between the midpoint of line $i$ ($L_i$) to the midpoint of line $j$ ($L_j$) [41]. Understanding how the cascading failure typically propagates through the network can help characterize real cascading patterns and provides methods for model validation.

$$\lambda_k = \frac{z_k}{z_{k-1}}, \qquad k = 1, 2, 3, \ldots \qquad (4)$$

### C. Applications of Dynamic Models

This subsection provides a review in the field of dynamic analysis of cascading failures. It is worth noting that this review focuses on dynamic models that can fully or partially capture time-based cascading phenomena, rather than QSS models that consider simplified dynamic processes. Readers are referred to [7], [19], [20] for reviews which consider QSS models.

The response of the power system involves a wide range of dynamic phenomena on different associated timescales [43]. The broad timescales of the typical dynamics involved in cascading failures in shown in Fig. 1. Various dynamic phenomena involved in cascading failures are shown in boxes with their distinct operating timescales, including the inertia response, frequency control and voltage regulation of synchronous generator, power flow re-dispatch and demand variation. During a generation re-dispatch process, discrete cascading events can be triggered by protection relays due to violations of line rating, frequency, or voltage limits. The fact that the sequence of cascading events is closely related to the distinct timescales of dynamic phenomena has a significant impact on the modelling requirement of system elements within the timescale of interest. According to various research objectives, researchers generally categorize system dynamics into three timescales [4], [10]:

1) *Short-term dynamics:* including emergency control actions and lasting for seconds.
2) *Mid-term dynamics:* including cascading overloads and generation re-dispatch process and lasting for minutes.
3) *Long-term dynamics:* including daily demand variation and actions taken by system operators.

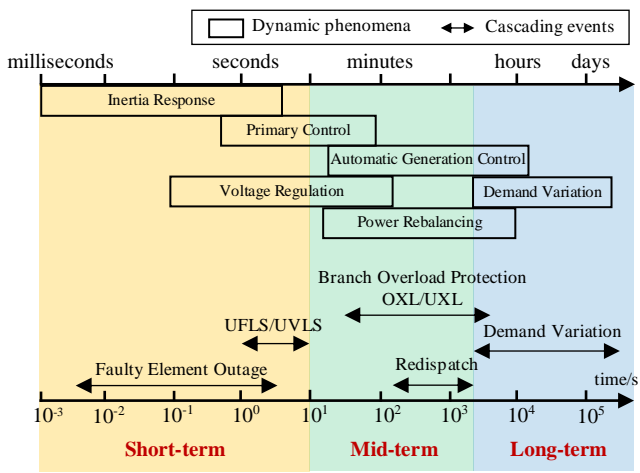

Fig. 1: Timescales of typical dynamics involved in cascading failures. Modified based on [10].

The use of time-based dynamic models has mainly focused on transient system behaviours following cascading events with simulators that are able to reflect the short-term and mid-term dynamics. Examples include ASSESS [44], POM-PCM [45], COSMIC [8], TRELSS [9], and other simulators implemented on available power system software programs, e.g., based on PowerWorld [8], Power System Analysis Toolbox [37] and DIgSILENT PowerFactory [46].

The short-term effects of cascading failures are evaluated by simulating large sets of potential disturbances (mainly N−k contingencies) that may occur during this timescale. Progress has been made in the risk assessment of cascading failures considering the probability and impact of cascading events, from which important properties observed from historical blackouts, such as self-organized criticality [16], power law distribution of demand loss [16], cascading propagation profiles [40] and other statistical/topological features [47], can also be effectively reflected. Moreover, dynamic models are frequently used for model comparison and cross-validation. Comparisons with QSS models show that QSS and dynamic models show high agreement at the initial stages of the cascades, but often diverge at the later stages when the dynamic instability drives the sequence of cascading events [8], [37], [48]. In QSS models, maintaining a generation-demand balance without overloading is one possible way to stop a cascade. However, frequency instability revealed in a dynamic simulator may undermine this assumption, leading to a more severe cascading. Furthermore, in the short-term and mid-term processes, the time allowed for system operator to correct for inappropriate control actions is often quite limited. This underscores the importance of data and analytical accuracy in assessing of short-term or mid-term dynamics.

There has been progress in simulating cascading failures with long-term state changes and system upgrades [49], [50]. Tools capable of analysing long-term effects typically decompose the short-term, mid-term and long-term dynamics into different phases. When simulating short-term dynamics, it is assumed that the states and parameters of long-term dynamics are constant. In [10], a quasi-dynamic cascading failure model is developed which deploys detailed dynamic modelling of short-term processes and takes into account the slow drivers of cascading failures such as demand variation. In operational planning, there are uncertainties associated with system operating states and the probability of cascading failures, but also a better scope of control strategies available for risk management. Preventive measures may include paying for certain generators as backup and dispatching additional generation reserves. The long-term analysis is of great value for the future development of power systems and for decision-making on mitigation strategies for dynamic cascading phenomena.

### D. Challenges and Possible Solutions

The main challenges to overcome in dynamic modelling of cascading failures are the computational burden and the large but often unavailable data required to describe dynamic characteristics of the main components that can affect cascade propagation. The mechanisms involved include re-dispatch capabili-ties of synchronous generators, frequency control, voltage regulation, demand management capabilities, available ancillary services, operator actions, etc.

Since cascading failure is inherently a large-scale power system phenomenon, detailed dynamic simulation of cascading processes contains thousands of state variables to represent system behaviour and relay actions. It, therefore, incurs very high computational costs. Considering the uncertainties introduced by the varying system states and various combinations of initial disturbances, it is very often infeasible to enumerate all possibilities in any practical-scale power system. To increase the simulation speed, various techniques have been developed:

1) To improve the solving speed of DAEs in dynamic power system simulation, and
2) To develop contingency screening techniques.

Khaitan and McCalley found that a multifrontal method solves DAEs an order of magnitude faster than the Gaussian elimination method [51], enhancing the efficiency of individual cascading failure simulations. Different screening techniques can support the efficient sampling of initial contingencies, including variance reduction in Monte Carlo simulation [52], selection of high-impact N−k contingencies [53], and fast sorting of contingency probability [54]. By applying contingency screening technologies, analytical tools can effectively model various types of disturbances and decide which features of cascading failures are highlighted and what conclusions can be drawn from specific analyses.

The network data and dynamic data of power system test cases, as the basis of cascading failure simulation, inevitably contain certain inaccuracies. As a result, validation against real blackout data is important to understand limitations and enhance the confidence of simulation tools. However, complete blackout datasets are rare, due to a lack of systematic data collection or confidentiality issues, although publicly available databases on outage data can be helpful in deriving conclusions on the cascading and blackout mechanisms [55]. Few simulators can be compared with a particular historical cascade. Although the tuning of simulators to reproduce past blackouts is important, it remains an open question whether such tuning will enhance the simulator's ability to provide useful decisions for other cascading blackouts [19]. Thus, instead of comparing with the sequence of cascading events, comparing the statistical properties of simulated results with historical records is another method of validation [39], [40], [42], [56]. The ability to capture important cascading features (e.g., power law distribution of demand loss) is considered as a positive indicator to support the validation of the simulator.

## IV. Case Study Application Using A Dynamic Cascading Failure Simulator

The application of a dynamic cascading failure model is illustrated in this paper using two representative test systems: the Illinois 200-bus synthetic system (ACTIVSg200) and the Texas 2000-bus synthetic system (ACTIVSg2000). Section IV-A describes the adopted dynamic model. In Section IV-B, a sensitivity analysis is performed on internal parameters (i.e., the number of failure scenarios and simulation step size) to establish the impacts that these features have on simulation efficiency and result accuracy. Following this, the dynamic model is compared with a QSS DC power flow model for cross-validation and to demonstrate the insights gained by using time-based dynamic simulation.

### A. Dynamic Simulator

The dynamic cascading failure simulator used in this paper has been developed by the authors and detailed in previous publications, with descriptions and source codes available in [46]. For brevity, only key aspects of the simulator implementations are described here. The dynamic model was developed based on DIgSILENT PowerFactory 2020 SP1 and MATLAB version 9.4 (R2019a) via the Python application programming interface. It implements automatic model set-up of power system components, controllers and protection relays, and thus shows good scalability to be easily applied to large power system models, without significant effort of creating and setting the bespoke cascading mechanisms from scratch. The frequency dependence of system components is explicitly simulated in the dynamic simulator, including the frequency responses provided by synchronous generators and loads, as well as different protection relays, such as overload protection, UFLS, OFGT and generator out-of-step protection (see Section III-A for details). Other dynamic phenomena (e.g., voltage instability) can be added if desired without affecting the general analytical methods presented in following subsections.

### B. Analysis on Selection of Simulator Parameters

Here, the sensitivity of simulation accuracy to the number of failure scenarios and the size of simulation time step are evaluated. It is clear that a smaller time step and a larger set of failure scenario can provide more accurate results but takes longer to simulate. It is important to investigate how the parameters of a dynamic model can be manipulated to produce consistent and useful results, so that advances can be made on understanding the level of detail required to model different cascading mechanisms.

#### 1) Number of Failure Scenarios:

It is computationally infeasible to perform a complete N−2 contingency analysis in large-scale power systems. For example, when ignoring the sequence of the initial outages, a complete N−2 analysis in a 2000-bus system needs more than $2\times10^6$ simulations. Thus, this section investigates how the estimated EDNS would vary as the number of failure scenarios increases. Here, a time step of 0.01s is first arbitrarily considered (to ensure high result accuracy) and 2000 N−2 contingencies are randomly selected in both the 200 and 2000-bus systems.

Results are shown in Fig. 2, where the 2000 N−2 contingencies are randomly ordered and plotted in different curves to eliminate the effect of the simulation order of different N−2 contingencies on the results. It can be seen that in the 200-bus system, the estimated EDNS converges to a consistent value when the number of failure scenarios exceeds 1,000. In fact, in ACTIVSg200, the estimated EDNS increases by just 0.4% when the number of scenarios increases from 1,000 to 2,000. Therefore, 1,000 can be considered a reasonable sampling size for N−2 contingency analysis in ACTIVSg200. This will obviously halve the simulation time required. However, in the 2000-bus system, the estimation of cascading risk triggered by N−2 contingencies is far from converged after 2000 simulations, as the estimated EDNS continues to increase and there is some range in output seen. Hence, it is concluded that additional failure scenarios are required to produce valid risk estimation for this system.

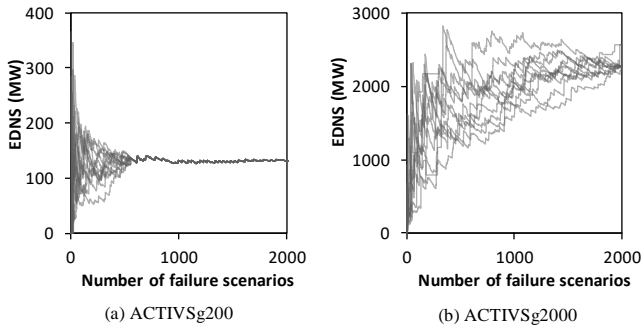

Fig. 2: Dependency of EDNS on number of failure scenarios using (a) ACTIVSg200 and (b) ACTIVSg2000 systems.

*2) Size of Time Step:*

Step size is another important factor in dynamic simulation. A larger step size can improve the calculation speed but may miss/delay the important cascading events that can alter the propagation of subsequent cascades. Here, 1,000 N−2 contingencies are performed on ACTIVSg200, with the step size increasing from 0.01 s to 0.3 s. The impact of increased step size on various cascading metrics is shown in Table II. It appears that the values of $VaR_{95}$, $CVaR_{95}$ and the average number of line outages are not sensitive to the increment of step size, and remain at a relatively constant level. However, the value of EDNS is more sensitive to step-size adjustments. It remains nearly constant for step sizes ranging from 0.01 s to 0.1 s, beyond which the estimated EDNS starts to decrease. This is because that the increased step size masks some of the cascading events (e.g., UFLS) that occur during this time interval, leading to errors in risk estimates. These errors can affect the mean unserved demand but have less impact on the extreme severe conditions (where VaR and CVaR are concerned) and the propagation of cascading outages.

To investigate root cause of these errors, CCDFs of demand loss with various step sizes are computed, as shown in Fig. 3. The results are the same for step sizes of 0.01 s and 0.05 s and then start to diverge as step size exceeds 0.1 s. When the step size is larger than 0.1 s, the dynamic simulator can still capture the large-scale (nearly complete) blackouts, as all curves converge to the same probability level when the unserved demand approaches 1.48 GW (i.e., the total demand in ACTIVSg200). However, it fails to capture some of the small-scale and mid-scale blackouts due to the large time interval between each sampling time point, leading to the reduction in estimated EDNS. In this case, a step size of 0.1 s is considered as an optimal selection to make a trade-off between simulation accuracy and efficiency. By increasing step size from 0.01 s to 0.1 s, 37% of the simulation time is saved, which ultimately takes 14.8 hours to simulate 1,000 N−2 contingencies on the 200-bus system, using a desktop PC with Intel Core W-2123, 3.60 GHz CPU and 32 GB RAM. It must be noted that this step size suggestion applies for this simulator which includes frequency dynamics and limited RES penetration. With voltage dynamics, and lower inertia, smaller step sizes will likely be needed. This is a topic of ongoing research.

TABLE II
IMPACT OF SIMULATION STEP SIZE ON CASCADING METRICS

| Time step (s) | 0.01 | 0.05 | 0.1 | 0.2 | 0.3 |
|---|---|---|---|---|---|
| EDNS (MW) | 47.37 | 47.37 | 47.35 | 45.32 | 39.54 |
| $VaR_{95}$ (MW) | 247.59 | 247.59 | 245.97 | 249.23 | 247.34 |
| $CVaR_{95}$ (MW) | 416.72 | 416.72 | 417.52 | 418.32 | 414.23 |
| Averaged line outages | 2.03 | 2.03 | 2.11 | 2.06 | 2.05 |
| Simulation time (hours) | 20.28 | 16.8 | 14.8 | 13.3 | 12.5 |

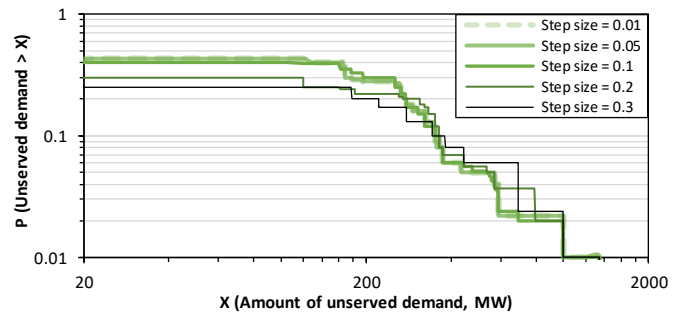

Fig. 3: CCDFs of the demand loss with increased step sizes, using the ACTIVSg200 system.

*C. Comparison with a QSS DC Power Flow Simulator*

A QSS DC power flow model (herein referred to as a *static* model) is adopted from [57], using DC optimal power flow (DC-OPF) solver from MATPOWER [58]. It starts with a DC-OPF solution that initializes the system state, and then computes cascading events iteratively until no further overload occurs or no active island exists. It can capture cascading phenomena such as thermal overload, re-dispatch capabilities of generation, UFLS and OFGT. This subsection demonstrates the similarities and differences between the dynamic and static models to provide useful insights into situations where it may be preferable to use a detailed dynamic model rather than the static model.

1,000 N−2 contingencies are simulated on ACTIVSg200 (with a step size of 0.1 s for the dynamic simulator as discussed in the previous subsection). Results are shown in Fig. 4. The static model tends to underestimate the small-scale demand losses because it fails to capture transient system dynamics following the disturbances, which often cause local frequency deviations and inter-area oscillation issues, and result in more load shedding. Moreover, the static model tends to produce longer cascades, in terms of the number of line outages, than the dynamic simulator. This is due to the fact that the static model requires to disconnect all the overloaded lines in one iteration, whereas the dynamic model disconnects overloaded lines is an inverse-time manner. Specifically, in the dynamic model and actual system operation, lines with higher overloads will be disconnected more quickly. Following this, generation will be re-dispatched and control actions, such as load shedding and generation disconnection, may occur, thus relieving stress on other transmission lines and reducing the number of affected lines.

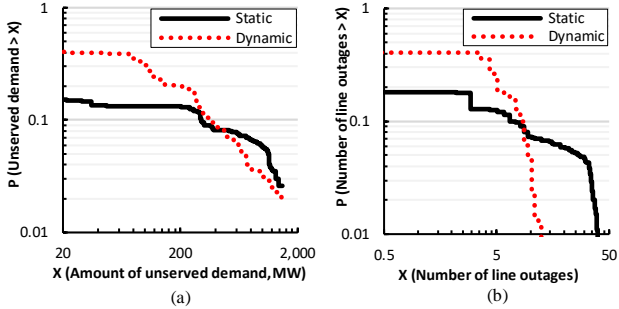

Fig. 4: CCDFs of (a) the unserved demand and (b) the number of line outages for static and dynamic models using ACTIVSg200.

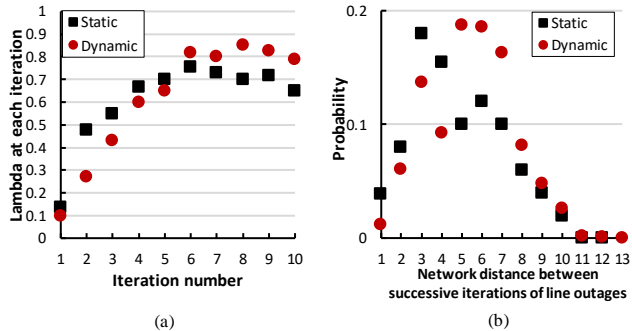

Fig. 5: The (a) propagation $\lambda_k$ and (b) probability density functions of network distance $d^{bus}(L_i, L_j)$ for static (squares) and dynamic (circles) models using ACTIVSg200.

Considering that system transient behaviours and relay actions are usually completed within 1 minute [40], successive outages with a time interval less than 1 minute are grouped into one iteration in the dynamic model. The static model inherently is an iterative algorithm. The process of generation re-dispatch and line overload check is iterated, and the number of line outages is recorded for each iteration. Based on this, the propagation $\lambda_k$ is computed and shown in Fig. 5 (a). In the results using both simulators, as the cascade progresses, $\lambda_k$ steadily increases until the 6th iteration, beyond which the data becomes noisy and tends to level off. The increase in $\lambda_k$ indicates an acceleration of cascading outages and a higher probability of further outages. The acceleration during the initial propagation phase of cascading failures is widely reflected in historical data and existing studies [40], [59], which highlights the validity of the cascading failure simulators. In addition, the static model provides higher values of $\lambda_k$ than the dynamic model during the initial phase, and the quickest failure propagation occurs at the 6th iteration. This emphasizes the important role of overload disconnection schemes in cascading failure modelling. The static model disconnects all the overloaded lines in each iteration, which can lead to rapid failure propagation and network separation. As a result, higher $\lambda_k$ values are observed during initial propagation using the static model. It is worth noting that the method of grouping line outages into iterations is different for static and dynamic models, but the steady increase in $\lambda_k$ supports model validation and provides insights into the impact of modelling assumptions on failure propagation analysis.

The probability density functions of network distance $d^{bus}(L_i, L_j)$ are computed based on the chains of cascading events and shown in Fig. 5 (b). The static and dynamic models show consistency in long-term cascading interactions but show divergence in shorter-term cascading interactions. More specifically, compared to the dynamic model, the static model shows more interactions at distances 3 and 4, and fewer interactions at distances 5-7. A particular benefit of this comparison between static and dynamic models is that it brings up areas of simulation that need to be adjusted to improve the match. Here, the results suggest that short-term cascading interactions are not adequately reflected in the static model. Further work is needed to improve static representations to better capture the dynamic cascading phenomena.

V. CONCLUSION

This paper discussed the state-of-the-art and future trends with respect to the dynamic analysis of cascading failures in power systems. The need for a time-based dynamic cascading failure model in evolving power systems has been emphasized and the advances needed to bridge the gap have been identified and discussed. The benefits and challenges posed by dynamic simulation of cascading failures in future power systems are outlined. Benefits include the ability to capture temporal characteristics and transient system behaviours following the disturbances. This has proven to have a significant impact on the risk estimation and mitigation of cascading failures, especially in future renewable-rich power systems. Stochastic renewable generation continues to replace traditional synchronous generators, resulting in low inertia issues and uncertain operating conditions. It therefore strengthens the need for a time-based dynamic model to incorporate complex system dynamics into future research and development. Issues related to dynamic simulation are mainly about the computational cost and the data availability for dynamic representation. Promising solutions to tackle these issues were discussed. Existing mathematical methods for fast DAE solving and contingency screening techniques can effectively alleviate the computational burden of dynamic models. Following this, applications of the dynamic simulator on the 200-bus and 2000-bus system showed that a trade-off between the simulation accuracy and efficiency can be achieved by adjusting simulation parameters. Finally, a comparative study between the dynamic model and a QSS DC power flow model showed that the QSS model tends to underestimate the small-scale outages and produces longer cascades with more line outages. The detailed dynamic model can be time-consuming for large-scale statistical studies but developing a comprehensive model as a reference can support the validation of different model assumptions, and facilitate the development of mitigation techniques, which requires detailed system dynamics to thoroughly test their capabilities.

ACKNOWLEDGMENT

The author acknowledges financial support from EPSRC (EP/L016141/1) through the Power Networks Centre for Doctoral Training.